# Decoding Orbital Angular Momentum in Turbid Tissue-like Scattering Medium via Fourier-Domain Deep Learning

**Avraham Yosovich[1], Anton Sdobnov[2], Alexander Doronin[3], Alexander Bykov[2], Igor Meglinski[4] and Zeev Zalevsky[1]**

[1] Faculty of Engineering and the Nanotechnology Center, Bar-Ilan University, Ramat-Gan, 5290002, Israel

[2] Optoelectronics and Measurement Techniques Unit, University of Oulu, Oulu 90570, Finland

[3] School of Engineering and Computer Science, Victoria University of Wellington, 6140 Wellington, New Zealand

[4] Aston Institute of Photonic Technologies, College of Engineering and Physical Sciences, Aston University, Birmingham, B4 7ET, UK

**Abstract**

Structured light beams carrying orbital angular momentum (OAM), such as Laguerre-Gaussian modes, are promising tools for high-capacity optical communications and advanced biomedical imaging. However, multiple scattering in turbid media distorts their phase and amplitude, complicating the retrieval of topological charge. We introduce VortexNet, a deep learning architecture that integrates an Angular Fourier Transform to explicitly extract rotational symmetries of OAM beams from experimentally acquired intensity and interference patterns. By transforming spatial information into the angular frequency domain, VortexNet isolates azimuthal features that persist despite scattering, enabling accurate topological charge classification even in complex optical environments. The results reveal that OAM-specific angular correlations can survive multiple scattering and be decoded through angular-domain learning. This establishes a new paradigm for structured-light analysis in complex medium, where deep learning enables the recovery of topological information beyond the reach of classical optics, paving the way for resilient photonic systems in communication, sensing, and imaging.

**Corresponding author**: Avraham Yosovich, yosovia@biu.ac.il

## Introduction

Light beams carrying orbital angular momentum (OAM), such as Laguerre-Gaussian (LG) modes, possess a helical phase structure characterized by a topological charge (TC) $\ell$ that quantifies the twist of the wavefront [1,2]. The ability to generate, manipulate, and detect these OAM states has opened new frontiers in optical communications [3], quantum information [4], and biomedical imaging [5], offering a means to encode information into the spatial structure of light [6].

Recent research has focused on the precise measurement and classification of the TC of LG beams, which is essential for the practical deployment of structured light in complex environments [5,7,8]. A variety of techniques have been developed to retrieve OAM states, including interference-based methods [9,10], diffraction through specially shaped apertures such as Y-shaped and triangular masks [11,12], and analysis based on intensity moments [13]. These methods allow for determination of both the magnitude and the sign of the topological charge. It has also been shown that beams with higher TC exhibit enhanced robustness when propagating through turbulent or scattering environments [14], and efforts have been made to manipulate TC, such as halving it using off-axis Gaussian beams [15], to enable flexible control over light's angular momentum.

Despite these advances, the practical deployment of OAM-based systems remains challenged by the effects of scattering in turbid medium [16,17], where beam coherence and phase structure degrade rapidly. Classical imaging techniques, such as interferometry, mode decomposition, diffraction-based analysis, and holography, often fail under strong scattering due to the randomization of optical fields caused by multiple scattering events [18–20]. These interactions disrupt phase coherence, spatial mode purity, and structured intensity profiles essential for accurate OAM retrieval [21,22]. The resulting speckle patterns erase the interpretable beam characteristics upon which conventional methods rely [23]. These limitations indicate that recovering OAM information in strongly scattering environments requires a fundamentally different, computational approach capable of extracting hidden topological signatures from distorted or apparently structureless optical fields.

Recent research has demonstrated substantial progress in preserving and recovering the orbital angular momentum (OAM) of light in scattering environments. Studies show that OAM-carrying beams can retain phase memory despite propagation through turbid media, exhibiting high sensitivity to refractive-index variations [24,25]. A range of novel techniques for OAM recovery has been reported, including ghost diffraction holography [26], scattering-matrix-assisted retrieval [27], and feedback-based wavefront shaping [28]. These approaches enable efficient demultiplexing of OAM channels, reducing crosstalk and improving data transmission under scattering conditions. In addition, non-interferometric methods have been developed to measure the modal composition of scattered helical beams [29], and the degradation of OAM under ballistic scattering has been systematically characterized [30]. Recovering OAM states is therefore critical for preserving information content, enabling quantitative sensing, enhancing imaging contrast, and supporting robust signal recovery in environments where structured light interacts with complex media. Collectively, these advances have significant implications for optical communications, biomedical imaging, and sensing technologies operating in strongly scattering or optically complex environments [31].

Recent demonstrations that OAM beams preserve their phase structure after propagation through multiply scattered tissue [24,25], identified as a key breakthrough in 2024 [32], suggest that topological information remains encoded within complex speckle patterns even when spatial coherence is lost. However, extracting this preserved angular information requires computational approaches capable of decoding OAM states from highly distorted, scattering-dominated optical fields.

Along this computational direction, Wang *et al.* demonstrated that deep learning can recover hidden topological information by reconstructing the full phase of multi-singularity SU(2) structured-light beams from minimal free-space intensity measurements, using a generative framework termed VortexNet [33]. While this work established the potential of neural networks for topological recovery under controlled, non-scattering conditions, it does not address the substantially more challenging regime of multiple scattering, where phase information and recognizable intensity structure are destroyed. In contrast, the present work introduces a conceptually different, Fourier-domain VortexNet designed specifically for the scattering regime, where only speckle patterns survive. By exploiting angular-frequency correlations rather than explicit phase information, our approach extends the VortexNet concept into a new physical domain, enabling, for the first time, reliable decoding of OAM states under strong multiple scattering.

Here, we introduce VortexNet, a deep-learning classification architecture, conceptually distinct from the generative phase-retrieval network introduced in [33], designed to decode OAM states from strongly scattered turbid medium where phase and recognizable intensity structure are lost, using angular-frequency analysis based on the Angular Fourier Transform (AFT) [34]. While recent studies have demonstrated CNN-based OAM classification in scattering environments [35–38], VortexNet operates directly in the angular frequency domain, enabling explicit extraction of azimuthal features that persist despite multiple scattering events. Unlike conventional CNNs that operate solely in Cartesian space, VortexNet transforms spatial data into angular frequency space, allowing it to disentangle azimuthal features from radial noise and scattering artifacts. This angular-domain enhancement gives the model a unique capacity to explicitly leverage rotational symmetries and azimuthal patterns characteristic of OAM and recover TC information even when spatial coherence is lost due to multiple scattering. This represents a conceptual shift in structured-light analysis, transforming how deep learning models interpret OAM information embedded in complex optical fields.

**Results**
We demonstrate that high-order topological charges imprint recoverable angular signatures throughout multiply-scattered media, maintaining perfect classification accuracy even at extreme optical depths ($z/l^*$) where spatial coherence is completely lost (Fig.1). VortexNet achieves perfect classification of the highest-order OAM state ($\ell = 5$) across all optical depths from quasi-ballistic to fully diffusive regimes ($z/l^* = 0$ to 16), maintaining 100% accuracy even where conventional methods completely fail. Integrated Gradients (IG) analysis reveals that this robustness arises from dynamic feature adaptation: the network progressively shifts from detecting full-ring petal structures in low scattering, to persistent radial spokes during annular-blur transitions, to subtle speckle harmonics in the vortex-memory regime, and finally to compact residual lobes under extreme diffusion. This adaptive strategy tracks physically meaningful OAM signatures throughout multiply-scattered media, demonstrating that high-order topological charges retain computationally recoverable angular information even when spatial coherence is completely lost.

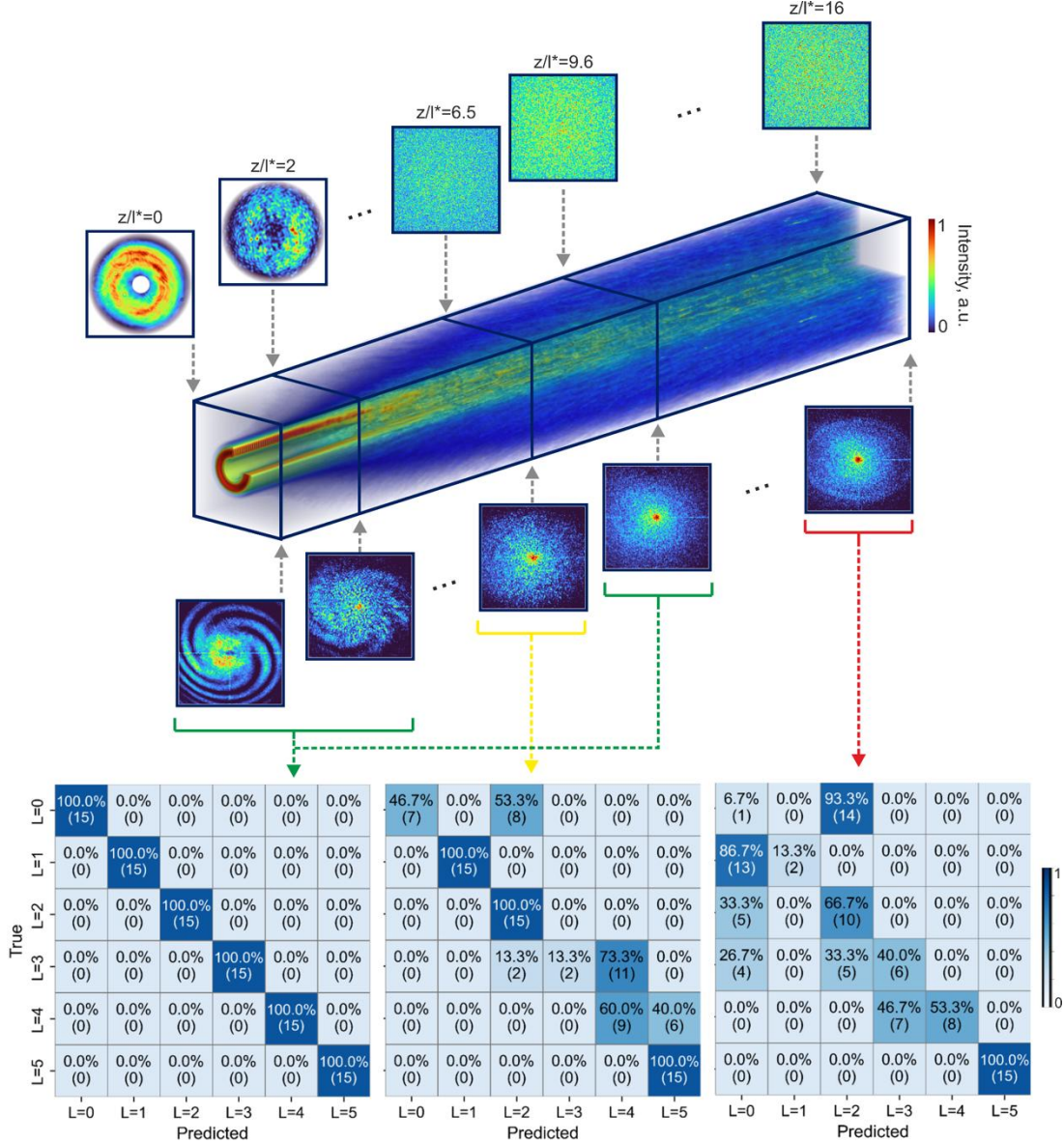

**Figure 1:** Evolution of OAM LG beam ($\ell = 5$) structure through multiply-scattered media obtained by Monte Carlo modeling [25], together with the corresponding depth-dependent classification performance. Top: Experimentally measured intensity distributions showing progression from the quasi-ballistic regime ($z/l^* \leq 4$, coherent donut profiles), through the annular-blur regime ($z/l^* = 6.5$, halo formation) and the vortex-memory regime ($z/l^* = 9.6$, dense speckle with preserved angular correlations), to the fully diffusive regime ($z/l^* = 16$, randomized fields). The corresponding on-axis interference (petal) patterns show the gradual degradation of the canonical five-lobed structure with increasing scattering strength. Bottom: Normalised confusion matrices at representative optical depths, revealing the classification performance of VortexNet.

Focusing on the $\ell = 5$ class in the Petals channel, the IG analysis reveals how VortexNet dynamically adjusts its feature attribution strategy across optical depths to maintain perfect classification accuracy. The confusion matrices (see Fig.1) reveal four clearly distinguishable depth regimes. In the quasi-ballistic regime ($z/l^* \leq 4$), IG clearly highlights the five-lobed petal structure and central phase singularity—canonical angular patterns associated with $\ell = 5$. These distinct interference features align with the model's 100% accuracy in this low-scattering region. As scattering increases into the annular-blur regime (z/l* ≈ 6.5), the outer petal fringes begin to degrade for many classes. However, for ℓ = 5, IG still emphasizes thin radial spokes radiating from the core, even as the ring structure is partially obscured. These persistent rib-like features allow the network to preserve its performance where lower-order states begin to fail. This dip arises from residual forward-scattered light that suppresses speckle contrast, so class-specific angular cues are washed out before fully diffusive statistics emerge. In the vortex-memory regime (z/l* ≈ 9.6), where the beam profile is dominated by dense speckle, IG uncovers subtle star-shaped lobes near the core—speckle harmonics that retain angular information due to the vortex memory effect. These faint yet physically meaningful signatures enable the network to continue identifying ℓ = 5 with full accuracy, even in highly randomized conditions. Finally, in the fully diffusive regime (z/l* = 16), where most structured light information is thought to be lost, IG reveals a compact core containing minute residual lobes. These remnants of the original five-fold phase symmetry remain detectable, allowing VortexNet to maintain perfect recognition of ℓ = 5, even as other classes collapse under extreme scattering.

The performance results, presented in Table 1, show that VortexNet maintains perfect accuracy across moderate optical depths, experiences a dip in the annular-blur regime, recovers in the vortex-memory regime, once the halo has decayed and the fully developed speckle retains subtle ℓ-dependent angular correlations that the network has learned to exploit, and then degrades further in the fully diffusive regime. Remarkably, the highest topological charge (ℓ = 5) is classified correctly at every depth. By focusing on angular features that persist despite multiple scattering events, VortexNet demonstrates that OAM information is not entirely erased. Instead, the network identifies residual patterns linked to OAM states, even under conditions where classical methods fail due to the loss of coherent phase fronts. Training with random rotations and cropping improves the model's robustness; however, classification accuracy becomes depth-dependent and non-monotonic. This variation reflects a balance between halo blurring (z/l* = 6.5), vortex-memory recovery (z/l* = 9.6), and the eventual wash-out of spatial correlations (z/l* = 16).

**Table 1.** Per-class and overall test accuracies (%).

| z/l* | Overall | ℓ = 0 | ℓ =1 | ℓ=2 | ℓ=3 | ℓ=4 | ℓ=5 |
|---|---|---|---|---|---|---|---|
| 0 (air) | 100 | 100 | 100 | 100 | 100 | 100 | 100 |
| 0.8 | 100 | 100 | 100 | 100 | 100 | 100 | 100 |
| 1.2 | 100 | 100 | 100 | 100 | 100 | 100 | 100 |
| 2 | 100 | 100 | 100 | 100 | 100 | 100 | 100 |
| 4 | 100 | 100 | 100 | 100 | 100 | 100 | 100 |
| 6.5 | 70 | 46.7 | 100 | 100 | 13.3 | 60.0 | 100 |
| 9.6 | 100 | 100 | 100 | 100 | 100 | 100 | 100 |
| 16 | 47 | 6.7 | 13.3 | 66.7 | 40.0 | 53.3 | 100 |

The results show that OAM state classification remains accurate over a wide range of optical depths, with particularly strong performance in the so-called vortex-memory regime. The model's accuracy peaks in the quasi-ballistic and vortex-memory regimes, dips at intermediate optical depths where low- and high-order features interfere, and declines under extreme diffusion where all structured information is lost. This non-monotonic trend mirrors the underlying physics of OAM propagation from coherent ring patterns, through blurred halos, to hidden speckle correlations, and ultimately to complete angular decoherence. The approach opens new pathways for robust structured-light applications in optical communications, topological photonics, biomedical imaging, and beyond.

**Discussion**

Our findings demonstrate that deep learning in the angular domain can uncover and exploit persistent OAM features in highly scattering environments, even when classical optics-based methods fail. VortexNet leverages the AFT to disentangle azimuthal features from spatial noise, enabling robust classification of topological charges across a wide range of optical depths. Remarkably, we observe a non-monotonic classification trend across scattering regimes. Accuracy is perfect in the quasi-ballistic regime ($z/l^* \leq 4$), dips in the annular-blur region ($z/l^* \approx 6.5$), rebounds in the vortex-memory regime ($z/l^* \approx 9.6$), and declines again in the fully diffusive regime ($z/l^* = 16$). This trajectory closely mirrors the underlying physics of OAM beam propagation: from coherent petal patterns, through blurred halos, to subtle speckle harmonics and eventual angular decoherence. The IG maps confirm that the model dynamically adapts its feature attribution, from rings, to spokes, to compact core structures, while consistently tracking physically meaningful OAM cues. The robustness of the $\ell = 5$ class across all optical depths is particularly striking. Even at extreme scattering ($z/l^* = 16$), where most beams become indistinguishable, $\ell = 5$ retains 100% accuracy. This suggests that high-order vortex modes imprint stronger angular signatures into the speckle field, supporting recent observations of phase memory in multiple-optical media [24,25]. VortexNet successfully exploits these residual correlations, offering a resilient path for recovering structured light information deep into turbid systems.

The obtained results have several major implications: (i) For optical communications, the approach opens the door to more robust OAM multiplexing in turbulent or diffusive channels. By enabling classification even when interference patterns degrade, deep learning could extend the usable range of OAM-based data transmission and reduce the dependence on coherent detection. While classification degrades beyond $z/l^* \approx 10$, hybrid schemes combining optical preprocessing with digital compensation may help preserve data integrity. (ii) In biomedical imaging, the ability to detect OAM states after deep tissue penetration can enable new modes of contrast and functional sensing. Since angular information persists beyond the reach of spatial coherence, VortexNet may facilitate high-specificity imaging even in challenging biological environments. The resurgence of accuracy around $z/l^* = 9.6$ suggests a "sweet spot" for probing scattering tissue using azimuthal harmonics. (iii) Compared to traditional methods, which rely on pristine interference or diffraction patterns, our model demonstrates clear advantages under realistic conditions. Classical techniques lose efficacy as soon as phase fronts scramble, while the data-driven nature of VortexNet allows learning from and adapting to distorted patterns, decoding OAM where theory alone would predict failure. However, limitations remain. The dataset is constrained in diversity and size, with only six topological charges considered. Dynamic or anisotropic scattering scenarios (e.g., flowing tissue) are not addressed. Expanding the dataset and incorporating time-varying elements will be necessary to generalize further. Additionally, integrating VortexNet with optical enhancement methods such as wavefront shaping or adaptive optics may boost performance at greater optical depths.

In summary, current study reveals that angular-domain deep learning can uncover hidden rotational order in highly scattered optical fields. VortexNet proves that OAM features are not irretrievably lost in complex media, they merely become encoded in forms that conventional optics cannot resolve, but which neural networks can learn to decode. This insight may help reframe how structured light is utilized in scattering environments, enabling a new generation of robust, data-driven photonic systems.

## Materials and Methods

### *Experimental Setup*

The experimental arrangement employs a modified Mach–Zehnder interferometer designed to explore OAM-carrying LG beams under controlled scattering conditions. This setup ensures that the input beam is well-defined, enabling a clean starting point to track the alterations induced by multiple scattering.

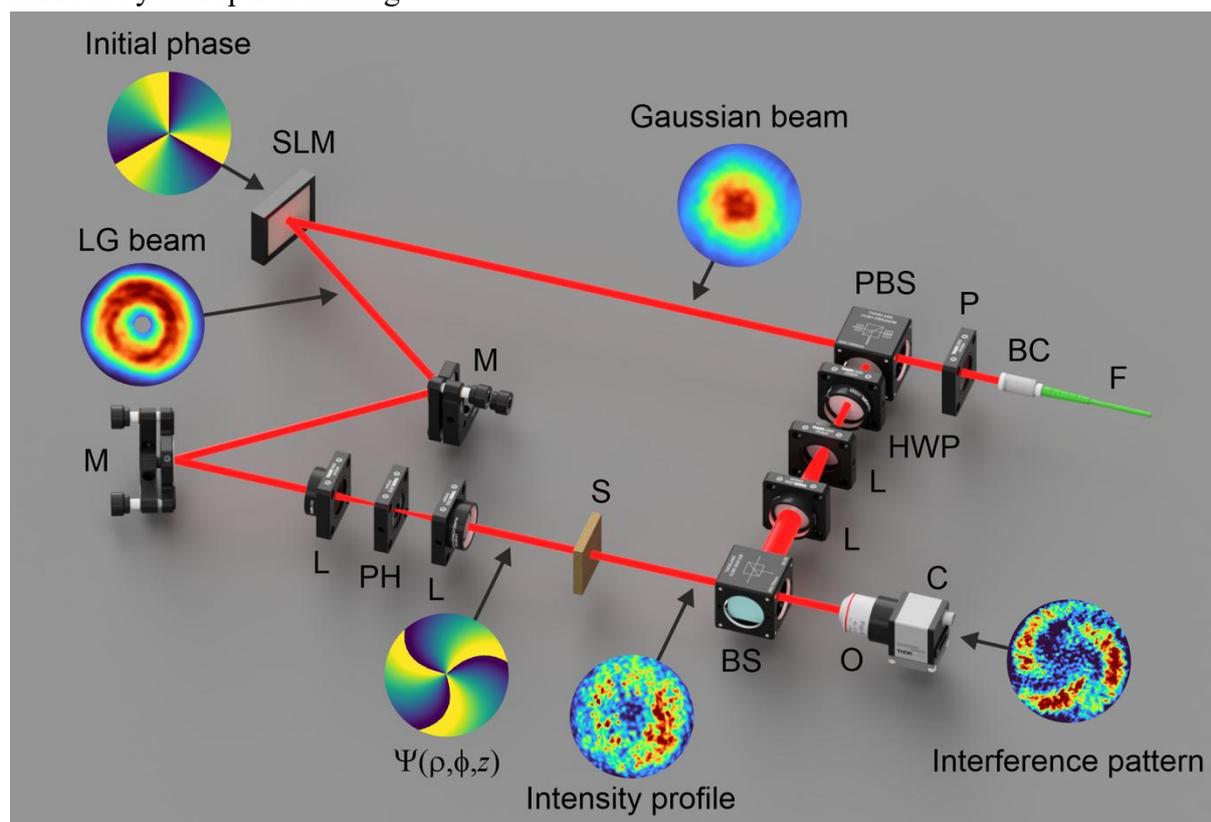

**Figure 2:** Schematic of the experimental setup: F– optical fiber delivering laser light; BC – beam collimator; PBS – polarizing beam splitter; SLM – spatial light modulator; M – mirror; L – lense; PH – pinhole; S – sample; HWP – half-wave plate; BS – beam splitter; O – objective; C – camera. A detailed description of the optical setup is presented in the main article text.

The arrangement (Fig.2) is a modified Mach–Zehnder interferometer tailored for OAM studies. A 640 nm, 40 mW continuous-wave laser diode is first coupled into a single-mode fibre (FM) for spatial filtering and then collimated (BC) to a ≈ 1.6 mm waist. After, linear polarisation is set by a polariser (P), a polarising beam-splitter (PBS) divides the beam into two arms. In the sample arm, forked diffraction gratings projected on a spatial light modulator (SLM) generate the desired Laguerre–Gaussian modes. The first-order diffraction is isolated with a pinhole (PH), re-collimated, and passed through the scattering cuvette (S). The reference arm is expanded by lenses L1–L2, adjusted in polarisation with a half-wave plate (HWP), and recombined with the OAM beam at a non-polarising beam-splitter (BS). A neutral-density filter (NF) balances the arm intensities. Finally, an objective (O) images the on-axis petals, off-axis interference, and LG-intensity channels onto a 1936 × 1216-pixel CMOS camera. This configuration provides a clean, stable reference while capturing three complementary views of each beam for the VortexNet training set.

*Scattering Medium*

Following established phantom preparation methods [42,43], the tissue-mimicking samples with different optical depth $z/l^*$ were prepared (here, $z$ denotes the medium's thickness and/or the depth of the beam penetration, and $l^*$ is the transport mean free path, i.e., the average distance over which the beam's original propagation direction is randomized by scattering). The samples spanned regimes from low scattering ($z/l^*$= 0.8, 1.2, 2, 4, 6.5) to strong multiple-scattering ($z/l^*$= 9.6, 16). In the low-scattering regime, the beam retained recognizable spatial characteristics, whereas under multiple scattering the medium generated complex speckle fields that significantly distorted the original Laguerre–Gaussian beam structure. For the $z/l^* = 0$ condition, no sample was placed in the optical path, and the LG beam propagated through air.

*Data Acquisition*

For each $z/l^*$ value the next dataset was captured by camera: stack of 100 frames with LG beam intensity profiles (reference arm was closed in this case to prevent interference); stack of 100 frames with on-axis interference intensity profiles (in this case sample beam and reference beam are parallel which is results in appearance of the petals structure); stack of 100 frames with off-axis interference profiles (in this case sample beam and reference beam are not parallel which is results in appearance of interference lines). A detailed description of both on-axis and off-axis regimes for the proposed setup is provided in [24]. The resolution of each frame was 1936 × 1216 pixels, offering detailed spatial information.

*Data Preparation*

We combined three distinct image types: Petals (on-axis interference intensity profiles), Off-axis Interference intensity profiles, and LG Beam intensity profiles into single three-channel composite images. Each image was kept at the native 1936 × 1216 pixels and normalized by dividing pixel values by 255. mixed-precision training and gradient accumulation kept the GPU memory and runtime manageable without the need to down-sample.

*Monte Carlo modeling*

The Monte Carlo (MC) method is widely regarded as the gold standard for simulating light-tissue interactions in biomedical optics, biophotonics, and related fields [44]. Since its introduction in the early 1980s, the role of MC modelling in photon transport analysis has expanded substantially alongside advances in computational resources [45]. MC tracks so-called photon packets to estimate radiance, polarization and related quantities, achieving higher accuracy with an increasing number of photon packet trajectories [45]. Despite its large computational overheads, GPU-based computation has made MC highly efficient, empowering recent studies of spectra, polarization, coherent back-scattering, OAM, machin learning (ML) and other applications [25,44-49].

In this work, we extend and apply our newly developed energy-efficient MC method [50]. The *in-house* algorithm is optimized for Apple's ARM-based M-series processors using *Metal* framework, leveraging their low-power, high-performance architecture to accelerate photon transport simulations. Available online and released as an open-source package, the method is enhanced with AI-based sample reconstruction and is readily adaptable to a wide range of applications [49]. For OAM beam simulations, we configure the full 3D volumetric voxel environment, enabled for the first time by Apple's unified memory architecture, which allows tracking up to $10^8$ photon trajectories in a few minutes. Our MC method performance has been benchmarked against traditional biomedical imaging solvers, demonstrating comparable accuracy with significantly reduced computational time and energy consumption [49,50].

*Deep-Learning Model: VortexNet*

VortexNet is a custom CNN developed to classify the TC ($\ell$) of LG beams transmitted through scattering media. A key innovation of this model is the integration of a numerical AFT [34], which enables the extraction of subtle rotational symmetries characteristic of OAM beams. The AFT approach analyses signals along the angular coordinate in polar space. By representing the beam's spatial intensity and phase distributions in polar coordinates and performing a Fourier transform along the angular axis, AFT isolates angular frequency components that encode OAM-specific features. This transformation effectively decouples azimuthal dependencies from radial ones, allowing the network to learn discriminative patterns tied to the beam's helical phase structure. VortexNet processes three input image channels: (i) petal interference patterns, (ii) off-axis interference patterns, and (iii) LG beam intensity profiles. Each input is subjected to AFT, yielding complex-valued angular spectra. The real and imaginary components are separated and stacked, effectively doubling the number of channels and enriching the feature space with OAM-relevant information. Following the AFT-based preprocessing, the network architecture employs standard deep-learning components. Convolutional and residual layers progressively refine the angular features into higher-level representations [40], while global average pooling and fully connected layers project these features onto class probabilities. Dropout regularization is optionally applied to improve generalization and reduce overfitting. By embedding angular-domain analysis directly into the learning pipeline, VortexNet achieves robust classification of OAM states under challenging scattering conditions. Even when beam profiles are heavily distorted, the network's ability to extract and amplify persistent angular signatures enables accurate recovery of topological charge, demonstrating its relevance for applications in optical communications, quantum photonics, and biomedical imaging. Figure 3 schematically represents the VortexNet model.

By training on experimentally generated three-channel images capturing the LG beam intensity, on-axis interference (petal) patterns, and off-axis interference fringes (Fig.3), VortexNet learns depth-robust angular features that enable accurate topological-charge classification.

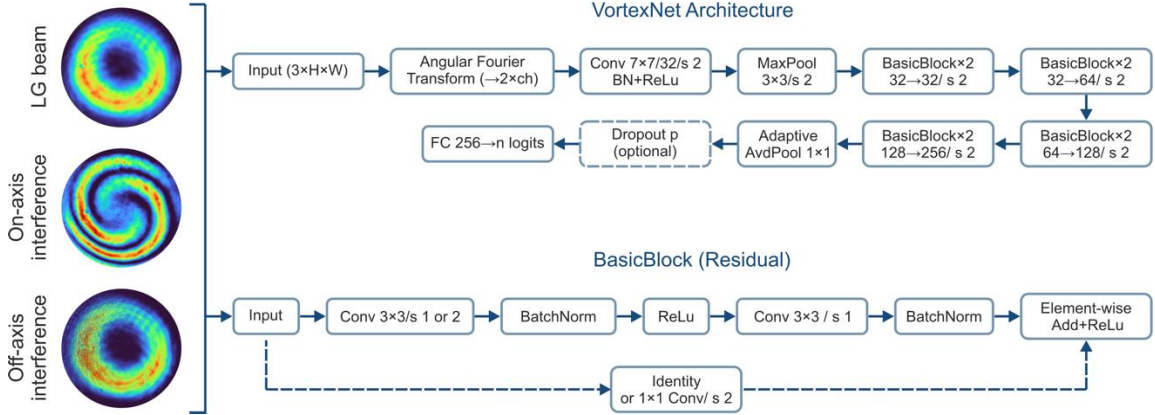

**Figure 3:** Architecture of VortexNet. The model ingests three speckle-derived input channels—on-axis interference (petal) patterns, off-axis interference fringes, and LG beam intensity distributions, which are first converted to polar coordinates and processed using AFT. The real and imaginary components of the AFT spectra are concatenated, yielding six feature channels. An initial $7 \times 7$ convolution followed by a $3 \times 3$ max-pooling operation reduces the spatial resolution, after which four residual stages progressively downsample the feature maps while increasing their depth. A global adaptive average pooling layer produces a 256-dimensional feature embedding, which is passed to an optional dropout layer ($p = 0.5$) and a fully connected classifier.

*Training Strategy*

Following preprocessing, we shuffled the list of 100-frame sequences while preserving the temporal order within each sequence. These were split chronologically: the first 80 frames (80%) were used for training, the next 5 frames (5%) for validation, and the final 15 frames (15%) for testing. This approach prevents data leakage between sets and ensures that model performance reflects true generalization rather than temporal redundancy. To enhance robustness, we applied on-the-fly data augmentation during training, including random rotations and spatial cropping. These augmentations increase the effective diversity of the training set and encourage the model to learn intrinsic angular features rather than overfitting to specific beam orientations. This strategy complements VortexNet's AFT-based architecture, promoting rotational invariance and improving generalization to previously unseen configurations. VortexNet was trained using the Adam optimizer [41] with a learning rate of 0.0001, a weight decay of 0.0001, and cross-entropy loss. We used a batch size of 32 with gradient accumulation over 4 steps. Dropout layers with a rate of 0.5 were incorporated to further reduce overfitting. Training proceeded for up to 50 epochs, with early stopping activated after 10 consecutive epochs without improvement in validation loss. We retained checkpoints for the five best-performing models based on validation accuracy, ultimately selecting the top model for final testing.

*Attribution analysis (Integrated Gradients)*

We used Integrated Gradients (IG) to attribute the class score $F_c(x)$ to input pixels [39]. IG was computed with a zero baseline using $m = 50$ steps (Captum), targeting the predicted class. Heatmaps show $|IG|$ log-scaled and 99th-percentile clipped, normalized to $[0, 1]$. Let $x \in \mathbb{R}^{C \times H \times W}$ denote the input image (three channels: Petals, Off-axis Interference, LG Beam intensity) and let $x_0$ be a baseline image (all zeros). For a differentiable scalar function $F_c(x): \mathbb{R}^{C \times H \times W} \to \mathbb{R}$ (the class-c score), we use the pre-softmax logit unless stated otherwise), The IG attribution for pixel i is defined along the straight-line path $\gamma(\alpha) = x_0 + \alpha(x - x_0), \alpha \in [0,1]$:

$$IG_i(x) = (x_i - x_{0,i}) \int_0^1 \frac{\partial F_c(\gamma(\alpha))}{\partial x_i} d\alpha \quad (1)$$

We approximate the path integral by a Riemann sum with m steps $\alpha_k = \frac{k}{m}, k = 1 \ldots m$:

$$IG_i(x) \approx (x_i - x_{0,i}) \frac{1}{m} \sum_{k=1}^{m} \left. \frac{\partial F_c(x_0 + \alpha(x - x_0))}{\partial x_i} \right|_{\alpha = \alpha_k} \quad (2)$$

We used PyTorch-Captum (IntegratedGradients) with a baseline of $x_0 = 0$ (an all-zero image), a target set to the predicted class for each image, $N_{steps} = 50$, and an internal batch size of 4 to manage memory.

For display, we use the absolute IG magnitude per channel, apply $\log(1 + \cdot)$, scaling, clip at the 99$^{th}$ percentile, normalize to [0,1], and render heatmaps with the Inferno colormap. overlays use transparency $\alpha = 0.65$. IG maps shown in the main text focus on the Petals channel (Fig. 1) to highlight azimuthal structure. Off-axis and LG-intensity channels show consistent trends and are available upon request. These maps were used to interpret depth-dependent cues (rings → spokes → compact core lobes) that track OAM features across scattering regimes.

**Disclosures**

The authors declare no conflicts of interest regarding the research described in this paper.

**Code and Data Availability**

The code and data supporting the findings of this study are available from the corresponding author (Bar-Ilan University) upon reasonable request. All data related to the experiments described in this article are archived on a lab computer at the University of Oulu. All data are available from the corresponding author upon reasonable request.

**Funding**

Authors acknowledge the support from the European Union's Horizon 2020 research and innovation programme under grant agreement No.863214 -- OPTIPATH project Pathfinder, European Innovation Council (EIC) and the European Cooperation in Science and Technology: Horizon 2020 COST Action CA23125 –- The mETamaterial foRmalism approach to recognize cAncer (TETRA) and COST Action CA21159 –- Understanding interaction light - biological surfaces: possibility for new electronic materials and devices (PhoBioS). In addition, I.M. also acknowledges support of British Council and the Department for Science, Innovation and Technology (DSIT), UK under grant project iPOL-Bio –"Integrating Polarized Light in AI-driven Biophotonics: Enhancing Health Applications". Authors also acknowledge the support from the UKKi UK-Israel innovation researcher mobility.